# Echocardiographic Image Quality Assessment Using Deep Neural Networks

Robert B. Labs[1], Massoud Zolgharni[1, 2], Jonathan P. Loo[1]
[1] School of Computing and Engineering, University of West London, London, UK
[2]National Heart and Lung Institute, Imperial College, London, United Kingdom

robbie.labs@uwl.ac.uk

**Abstract.** Echocardiography image quality assessment is not a trivial issue in transthoracic examination. As the in vivo examination of heart structures gained prominence in cardiac diagnosis, it has been affirmed that accurate diagnosis of the left ventricle functions is hugely dependent on the quality of echo images. Up till now, visual assessment of echo images is highly subjective and requires specific definition under clinical pathologies. While poor-quality images impair quantifications and diagnosis, the inherent variations in echocardiographic image quality standards indicates the complexity faced among different observers and provides apparent evidence for incoherent assessment under clinical trials, especially with less experienced cardiologists. In this research, our aim was to analyse and define specific quality attributes mostly discussed by experts and present a fully trained convolutional neural network model for assessing such quality features objectively. A total of 1,650 anonymized B-Mode images with dissimilar frame lengths were stratified from most popular ultrasound vendors equipment and clinical quality scores were provided for each echo cine by Cardiologists at England's Hammersmith Hospital which fed our multi-stream architecture model. The regression model assesses the quality features for depth-gain, chamber clarity, interventricular (on-Axis) orientation and foreshortening of the left ventricle. Four independent scores are thus displayed on each frame which compares against cardiologists' manually assigned scores to validate the degree of objective accuracy or its absolute errors. Absolute errors were found to be +/- 0.02 and +/- 0.12 for model and inter observer variability, respectively. We achieved a computation speed of 0.0095ms per frame on GeForce 970, with feasibility for 2D/3D real-time deployment. The research outcome establishes the modality for the objective standardization of 2D echocardiographic image quality and provides a consistent objective scoring mechanism for echo image reliability and diagnosis.

## 1 Introduction

### 1.1 Overview

During transthoracic (TTE) examinations, a quality image acquisition is a priced clinical enterprise mostly required for expert's assessment and quantification of cardiac functions. Due to its ubiquitousness and non-ionizing advantages, echocardiograms have found its significance in antenatal, obstetric and general diagnosis of cardiac infarction. Although, echo image does provide rich information about myocardium, it does not present crisp edges of a well-defined resolution when compared to photographic images. This inherent limitation in echo image resolutions poses a challenge to clinical measurement and interpretation of image features, the reason it is solely consider suitable for experts. In addition, echo image acquisition requires significant skill, absence of which further exacerbates reliability and image quality. To get around these problems, quality assessment is carried out as a control to prevent suboptimal image quality from being quantified. Nevertheless, the method of image quality assessment is a subjective process, where an echocardiography specialist visually inspects the images and decides on what anatomical features present in the image to be pathologically relevant. This process is laced with a spread spectrum of opinion and decision variability [1] even when an image is reassessed by the same operator. These variabilities and uncertainties [2] are found to impair quantification accuracy of cardiac functions, diagnosis and the overall quality of patience care.

A two-dimension (2D) echocardiographic images therefore, requires an objective assessment with a view of achieving optimum image quality, reproducibility, accurate quantifications and to provide platforms for automated diagnosis. Much more of a necessity is the requirement to define a set of mandatory attributes which constitutes an objective standard in quality assessment procedure. Unfortunately, varying consensus still abound among clinicians and equipment manufacturers on what element of 2D echocardiographic images constitutes relevance to standard quality [3] while Cardiologists' are saddled with task for quantifying cardiac functions and provide accurate interpretations within the confine of varying clinical scenarios.

To this end, this research investigated and propose novel method and approach to quality assessment using legacy and domain attributes. These are only paramount to the quantification of the left ventricle (LV) functions, which serves the ultimate purpose of cardiac diagnosis or myocardial assessment. This research employs the use of multi-stream deep convolutional neural network to extract image's quality attributes and computes objective quality score which can be used to guide operators in obtaining optimum image quality in real-time.

## 1.2 Related works

Prior to the use of deep convolutional models, several researchers proposed a series of methods for echo image quality assessment which were enhanced by the advent of deep convolutional neural networks. Some of the prominent research on real-time quality assessment using deep convolutional neural networks include four papers and our earlier research namely; [4], [5], [6],[7] and [8]. The authors in [6], [7] presented an algorithm based on cardiac view detection. The approach successfully models the detection of chambers in apical A4C echocardiography and admitted that the approach did not guarantee good performance when images consist of significant noise or low contrast pathologies are assessed. In the same vein, the author in [7] presented an algorithm based on convolutional neural networks to tackle cardiac view classification using a multi-class detection approach. The result yields a contradicting detection results on images with low contrast-gain and high contrast-gain which reinforce the conclusion on a model that combines spatial and temporal extraction to guarantee better classification accuracy.

To the best of our knowledge, the most recent work on automated quality assessment, is the work by Luong, C., et al., [8], Labs, R. et al., [7], .Dong, J. et al., [9] and Abdi et al.[10]. Abdi's work was based on a regression model to elicit the automatic scoring for five apical standard views. Quality scores were estimated based on the sequence of echo cine loops which include the end systole, end diastole to produce a single quality score per frame per view. The results were impressive with a prediction accuracy of 86% and a computation time of less than one second on a desktop computer. Unfortunately, Abdi's work used a weighted average of quality measure hence, the scores do not provide precise guidance to the aspect of image quality that needs to be optimized. On the other hand, Luong's [8] which investigated the mechanically ventilated TTE on hospitalised patients, yielded certain improvement in performance but with much larger dataset to represent wider population distribution. Luong's method of quality assessment was similar to Abdi's in the sense that the unified model produces a single score per image view across the nine apical standards considered. Luong admitted there exist no reference standard for the evaluation of echocardiographic image quality [8] but a scale of criterial used in many publications does not represent expert visual assessment and consensus on 2D echocardiographic image quality. Similarly, Dong [9] proposed a generic quality control framework on A4C. It considered application of image quality to fetal ultrasound to alleviate the challenges in antenatal investigation. The proposed method detailed the assessment of image quality using two features namely Gain and Zoom. It was considered as first comprehensive quality control system but significantly lacks adequacy for generalisation of quality attributes required for wider use case. Hence, suitability for quality assessment is inherently impaired.

In our study, we defined and modelled four quality attributes as this separately provide the most relevant quality assessment information for operators' feedback during the image acquisition and for generalised standard benchmarking purpose. This work is based on multi-stream regression model with selective qualitative attributes which are progressively distinguished from [10]. The advantage of this novel method of quality assessment provides the specific component of quality need to be optimised and guarantees clinical real-time feedback for optimum image quality in the lab. This means that during the acquisition phase, the operators can assess specific quality element independently, as would be indicated on each four attributes rather than obtaining a weighted average of quality components which is the existing and current assessment method obtainable in the most recent research papers.

To the author's knowledge, there are no published methods on attributes of quality and its assessment method in echocardiography modelling. Our novel approach and quality formulations can be used to assess, optimise and quantify echo images surgically, in real-time.

### 1.3 Main Contributions

Interpreting the results of the proposed architectures in the literature is not straightforward. This is because a direct comparison of the models' performance would require access to the same patient dataset. At present, no echocardiography dataset and the corresponding annotations for the image quality assessment is publicly available. We, therefore, aimed at evaluating the performance of deep learning models for the automated image quality assessment using an independent (PACS) echocardiography dataset which would be made available at IntSav repository.

In the view of the above, the main contributions of this research can be summarized as follows:

- Novel formulation of quality attributes for 2D echocardiographic images and novel method of its objective assessment
- Annotation of an independent echocardiography patient dataset showing four attributes of image quality namely: fore-shortening, chamber clarity, depth gain and axial orientation for A2C, A4C apical standard views.
- Public release of complete annotated patient dataset to allow future studies and external validation of the new approach or methods.
- Demonstrate the feasibility and applicability of four quality attributes framework which can be adapted for benchmarking, reference standard of evaluation and objective quality scoring of 2D echocardiographic cine loop.

## 2 Materials and Methods

### 2.1 Definition of Legacy Attributes of 2D Image Quality:

Most of the subjective criterial used for the assessment of image quality [4], [9], [10], [11], in echocardiography and implemented under point of care workflow, can be classified under legacy attributes and domain attributes. Hence, we classified chamber clarity, image depth gain, artefacts and probe resolution density under legacy attributes. In the study, an objective consideration for chamber detection/chamber clarity and depth-gain were explored and we proposed as follows:

(i) **Chamber Clarity** is a quality attribute defined by several distinguishable pixel's formation in the echo image to reveal the left ventricle (LV), right ventricle (RV), right atrium (RA) and left atrium (LA). This attribute is summed up using the root mean square RMS contrast, equation (1), the visual perception of the subjective element of quality in echocardiography. Contrast provides the perceptual ability to distinguish between luminance levels and very common in medical imaging, to have echo data with low-contrast or very high-contrast yet bearing anatomical structures required for quantifications and diagnosis. Typical examples of varying contrast levels are given in Fig 1. Low image contrast possesses a significant challenge where limited expertise is found. For cardiac images, the root mean squared (RMS) contrast; equation (1), $C_{i,j}$ which does not depend on angular frequency content or spatial distribution is best suited for 2D cardiac images and given by standard deviation of normalised pixel intensity $I_{i,j}$ for a given pathology, the area of interest; where *(i, j)* represents the *i-th* and *j-th* element of 2D image size M, N; in this case 227x227 used in our modelling. Contrast scores spans ranges from 0 – 9. 4.5 for the most obvious poor contrast, 6.0 for very high contrast and 9.0 for optimum contrast where relevant anatomical details are clearly visible. Chamber clarity is represented in our model as 'LC' attributes representing LV chamber clarity.

$$C_{i,j} = \sqrt{\frac{1}{MN}\sum_{i=0}^{N-1}\sum_{j=0}^{M-1}(I_{i,j} - \bar{I})^2} \qquad (1)$$

(ii) **Depth-Gain** constitute a prominent attribute of 2D echocardiography [6], classified under legacy attribute. Image gain for anatomical features within the first few centimetres in the image sector may become excessively high or the lower sector of images becomes excessively low and marred with artefact which potentially affects visibility or obscure relevant anatomical details as illustrated in Fig 2. This is peculiar to 2D echocardiography because of the way the ultrasound image is formed through acoustic frequency and propagated in trabeculated tissues. Improper depth-gain can induce significant lack of uniformity in the pixel intensities across the image especially in the lower part of the image sector. Since echocardiographic images are formed by reflected beams, they are susceptible to depth change, sector sizes, and pathological differences. Equation (2) describes the intensity of reflected beam, which is associated with depth gain; where $d^2\phi$ represent the luminous flux of the infinitesimal area of source $d\sum$, dividing by the product of $d_{\sum}$, infinitesimal solid angle $d\Omega_{\Sigma}$ and $\Theta_{\Sigma}$ angle between the normal $\Omega_{\Sigma}$ to the source $d\sum$. While luminance is the photometric measure of the pixel luminous intensity per unit area of light at a given area of interest, brightness therefore is the subjective impression of the object of luminance $I_{i,j}$ and is measured in candela per square meters cd/m$^2$. Since we are considering discrete signal samples in spatial domain, it unlikely that 2D echocardiographic image with varying luminous intensity can be referenced using Minkowski metric [12]. Therefore, photometric intensity, equation (3) as $\sigma_y^2$, the variance of *y* given a real valued sequence of *y* = {$y_{1....}$ $y_n$} with $\hat{y}$ as the mean of *y*. For optimum values of depth-gain, a time gain compensation (TGC) controls are often used to either decrease or increase in the near or far fields as appropriate to the area of interest in apical standard view. Our model learns from ground truth clinical scores and provides specific scores feedback to the operators depending on pathology under consideration. Depth-gain score is represented as 'DG' in our multi stream architecture.

$$I_{i,j} = \frac{d^2\Phi}{d\Sigma d\Omega_{\Sigma} cos\Theta_{\Sigma}} \qquad (2)$$

$$\sigma_y^2 = \frac{1}{n-1}\sum_{i=1}^{n}(y_i - \hat{y})^2 \qquad (3)$$

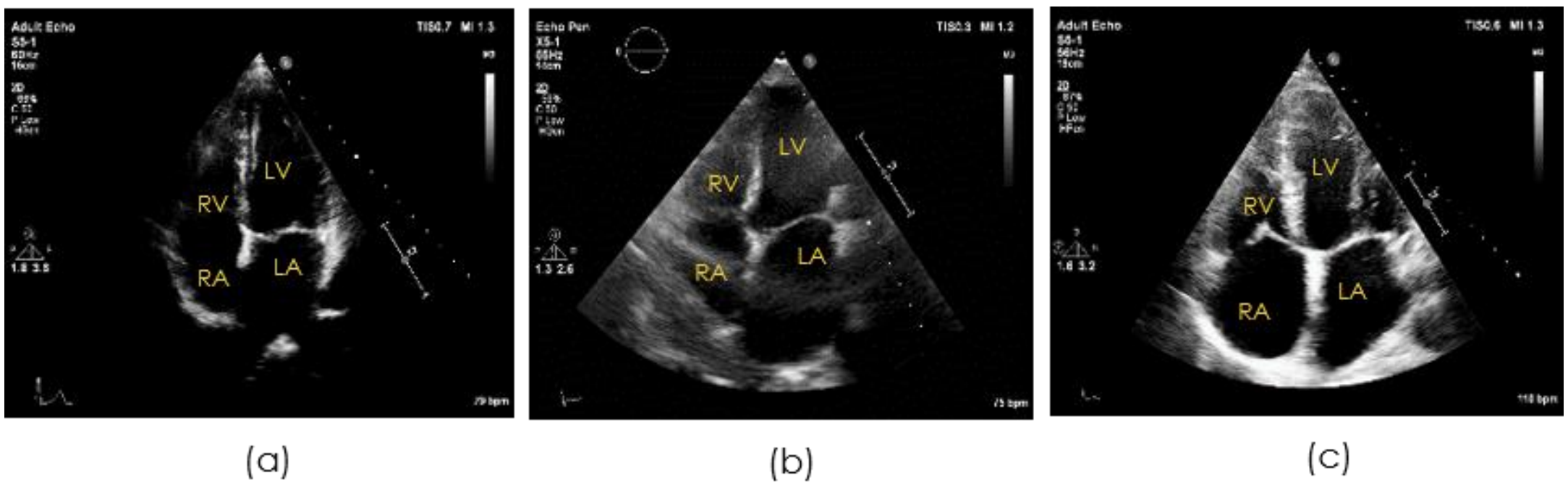

Fig. 1. Samples showing different assessment of chamber clarity of A4C. (a) chamber clarity in very low contrast image, (b) chamber clarity in average contrast image, (c) chamber clarity in optimum contrast images are all significant to diagnosis in TTE.

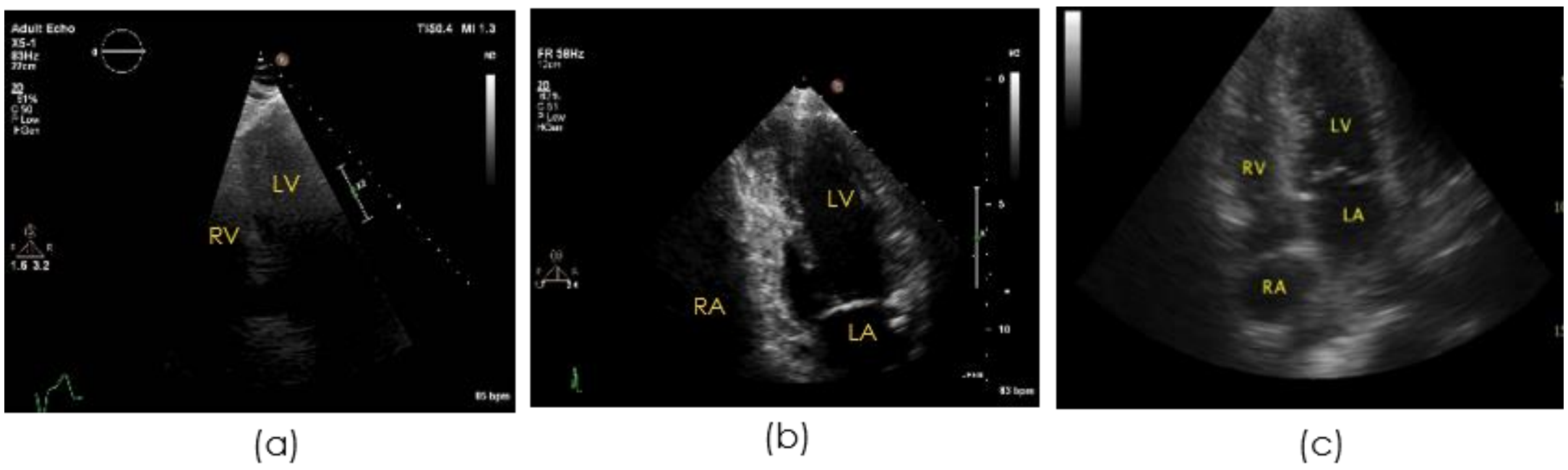

Fig. 2. Samples showing A4C's different assessment in image depth-gain; (a) poor depth-gain obscuring RV, RA & LA of the image sector. (b) inappropriate depth gain obscuring the RV & RA of the image sector, (c) showing four chambers in optimum depth gain. Its common in echocardiology to have a low depth gain images bearing significant anatomical details thereby making it relevant to cardiac diagnosis.

## 2.2 Definition of Domain Attributes of 2D Image Quality:

As with legacy attributes discussed, domain attributes of quality specify the imaging criteria which are essentially relevant and exist within cardiology domain [4]. We identified On-axis and Apical foreshortening as crucial attributes that contribute to quality assessment and functional quantifications. We explore these in our study and propose as follows:

(i) **On-Axis Attributes:** In apical four chamber (A4C) standard projection, identifying four heart chambers is crucial to functional assessment and accurate diagnosis. However, there is a complex interaction of the probe's beam cutting through the heart at specific angle, represent significant skill challenge because heart's anatomical features do not present crisp edges as a reason of trabeculated endocardium [8]. The acceptable projection ensures that the LV is positioned at the heart's apex and its visible during the systolic cycles while the interventricular septum orients vertically and divides the sector into two, running through the anterior and posterior plane to yield a four-chamber apical view as illustrated in Fig 3. On-Axis attributes is a function of probe's beam slicing through the centre of the apical chamber for optimum projection unless when the apex segment is cut off on purpose, which can be for a region of interest (ROI) analysis. We trained our model to recognise the degree of deviation and quantify the magnitude when ultrasound probe is medially or laterally translated which could cause image to be either off-axis significantly, mildly off-Axis or perfectly On-Axis.

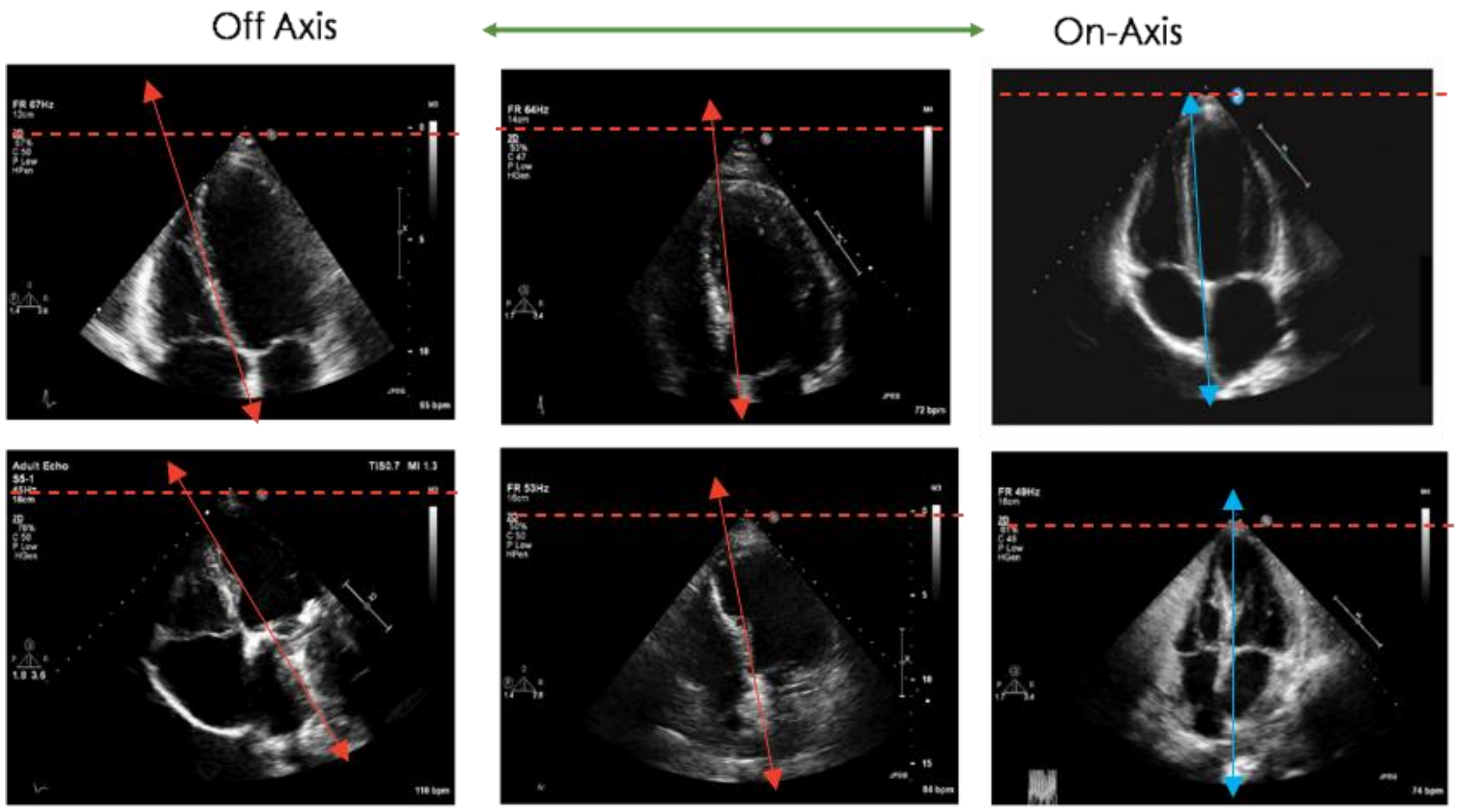

*(a)* *(b)* *(c)*

Fig. 3. Showing three samples, (a, b, c) of apical orientation: Significant Off-Axis image quality, Mildly off-axis image quality, On-axis image quality, respectively. An optimum on axis image quality shown the interventricular septum runs vertically down the middle of the screen indicated by blue arrows.

(ii) **Apical foreshortedness** is common in echocardiography workflow [14] and represent a significant attribute in image quality assessment. We explored the presence and the magnitude of foreshortedness as a domain attribute by which A4C image quality is assessed. Foreshortening describes the non-linear perspective transformation, a kind of structural deformation where changes in size of the areas of interest (AOI) and distances becomes geometrically incongruent. Fig. 4., illustrates apical foreshortening during systolic cycle and foreshortening examples in single frame study. Hence, the view projection is defined by many-to-one mapping where distances to the image is inversely proportional to image size resulting in scaling deformation, vanishing point and angular distortions. This deformation introduces a third plane to a 2D image, representing a significant distortion of planar figure, chamber volume becomes inaccurate [15] and extremely essential anatomical features that relates to quantification and diagnosis are obscured. Thus, foreshortening prevents the detection of crucial pathologies in the apical region, like supra-apical infarction, thrombus and reliable clinical measurements [8]. We refer to this transformation in terms of the product of homogenous properties of the N+1 image in equation (4).

$$\begin{bmatrix} 1 & 0 & 0 & 0 \\ 0 & 1 & 0 & 0 \\ 0 & 0 & 1 & 0 \\ 0 & 0 & -\frac{1}{d} & 1 \end{bmatrix} \begin{bmatrix} x \\ y \\ z \\ 1 \end{bmatrix} = \begin{bmatrix} x \\ y \\ z \\ -\frac{1}{d} \end{bmatrix} => \left( -d\,\frac{x}{z}, \quad d\,\frac{y}{z} \right) \tag{4}$$

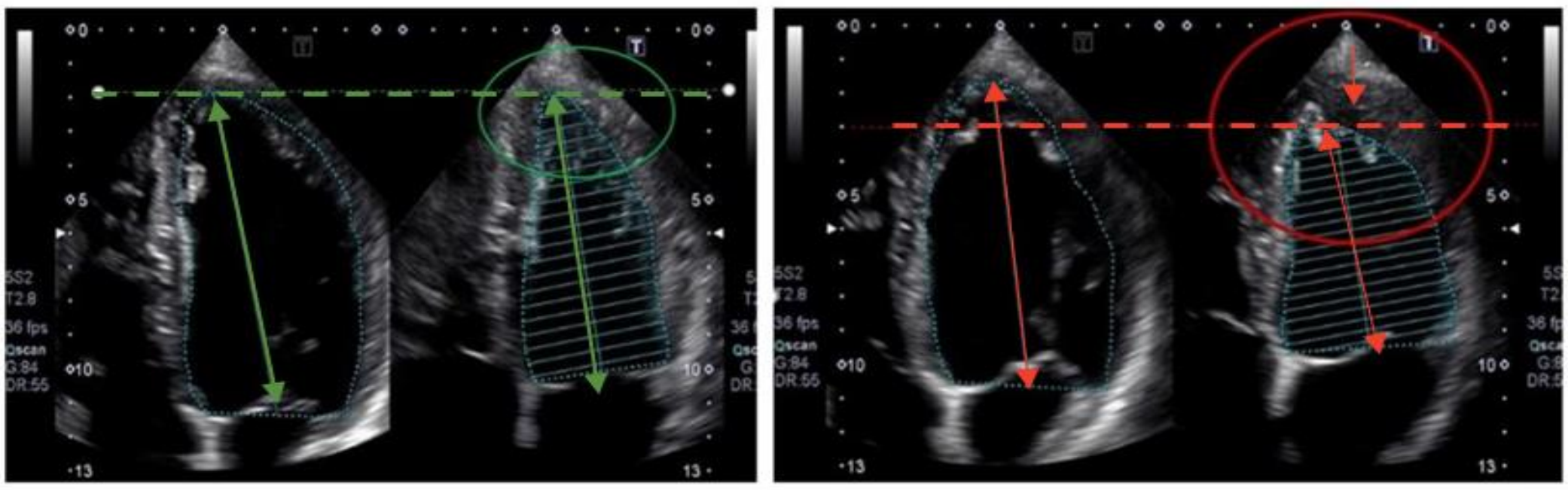

(a) (b)

Fig 4. Frame samples showing the effect of foreshortening in LV quantification. apex should remain the same during diastolic and systolic cycle (a) No foreshortening - the apex remains virtually the same during diastolic and systolic cycle, (b) Apex position alters during systolic cycle indicating apical foreshortening.

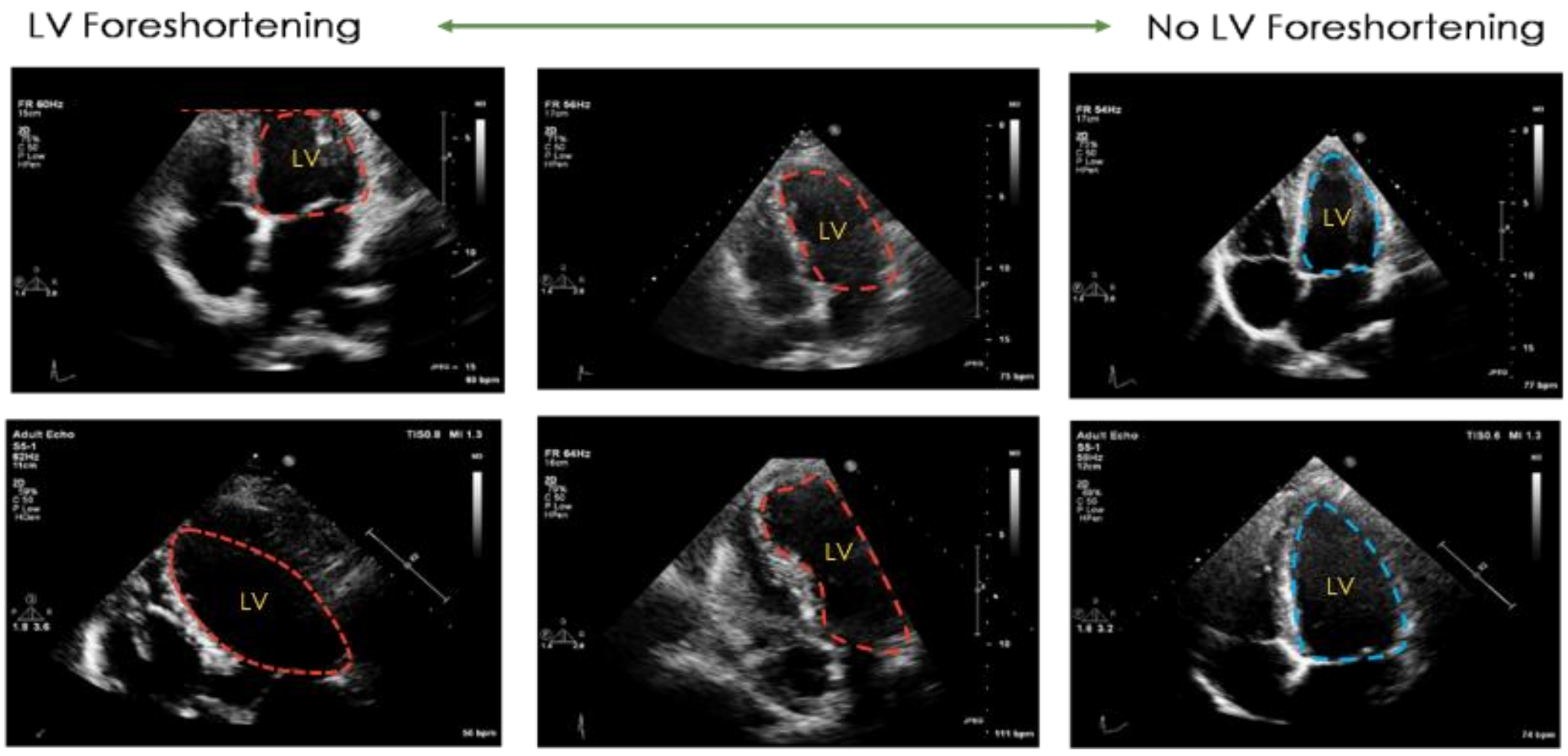

Severely Fore-Shorten Mildly Fore-Shorten Minimum Fore-Shorten

Fig 5. Our model is trained to recognize 3 levels of foreshortening: severe foreshortening, mild foreshortening and zero foreshortening attributes.

### 2.3 Ground Truth definition (Expert's Manual Score Criteria)

To model objective quality score for cardiac images, experts visually inspect each scan images using a developed interface that closely represent how the images are displayed during TTE in the laboratory. Two experts independently provided ground truth annotations using the criteria summarised in Table I. On each quality attributes annotation, an image with maximum score range of 4.0, 6.0, 9.0 are considered as low quality, average quality and optimum quality, respectively. These manual scores are represented as ground truth ($Q_{GTi}$) in our model.

TABLE I
MANUAL SCORES CRITERIA FOR ON-AXIS, CLARITY, DEPTHGAIN & FORESHORTEN ASSESSMENT GROUND TRUTH

| **Assessed Element** | **POOR QUALITY** | **AVERAGE QUALITY** | **OPTIMUM QUALITY** |
|---|---|---|---|
| Correct Cardiac Apex | 2.0 | 4.0 | 6.0 |
| Septum Visible | 1.0 | 1.5 | 2.0 |
| Interatrial Septum Visible | 1.0 | 1.0 | 1.0 |
| **Max. Score (On-Axis)** | **4.0** | **6.5** | **09** |
| | | | |
| 2/4 Chambers Clarity | 1.5 | 2.5 | 4.0 |
| Mitral Valve Clarity | 1.5 | 2.0 | 3.0 |
| Tricuspid Valve Clarity | 1.0 | 1.5 | 2.0 |
| **Max. Score (LV Clarity)** | **4.0** | **6.0** | **09** |
| | | | |
| Apex Signal Gain | 2.0 | 3.0 | 5.0 |
| Basal Signal Gain | 1.0 | 2.0 | 3.0 |
| No Excess Gain Artefacts | 1.0 | 1.0 | 1.0 |
| **Max. Score (Depth-Gain)** | **4.0** | **6.0** | **09** |
| | | | |
| LV Apex Visibility | **1.0** | **2.0** | **3.0** |
| Normal-Shaped Diastole | 1.5 | 2.0 | 3.0 |
| Normal-Shaped Systole | 1.5 | 2.0 | 3.0 |
| **Max. Score (F-shorten)** | **4.0** | **6.0** | **09** |

### 2.4 Data sources

The study population consisted of a random sample of (PACS2-Dataset) 1,039 Echocardiographic studies from patients with age ranges from 17 and 85 years, who were recruited from patients who had undergone echocardiography with Imperial College Healthcare NHS Trust. The acquisition of the images had been completed by experienced Echocardiographers using ultrasound equipment from GE healthcare (Vivid.1) and Philips Healthcare (iE33 xMatrix) manufacturers according to the standard protocols.

Ethical approval was obtained from the Health Regulatory Agency otherwise known as Integrated Research Application System (IRAS) identifier-243023. Patient automated anonymisation was performed to remove the patient-identifiable information. DICOM-formatted videos were then split into constituent frames, and 20 sequence frames were extracted from each echo cine loop while each frame bearing the same clinical score as its respective original echo cine score from each video to represent arbitrary stages of the heart cycle, resulting in 20,780 frames. The dataset randomly split into training (16,624 frames), validation (2,078 frames), and testing (2,078 frames) sub-datasets in a 60:20:20 ratio.

Consequently, the entire echo cine loops were independently studied between two cardiologists to create ground truth labelling annotation for ED/ES cycle and assigned clinical scores. In order to create an exclusive list for the 3 emerging classes, we implemented a visual aid software scorer in MATLAB, and allows each echo cine loop to playback the length of frames and using the score slider, a single score is assigned to the set of frames. We refer to the labelling as GT1 and GT2 from respective experts.

The disparity between the opinions was determined and found at 0.12 +/- 0.08. Thus, each clip belongs to one of the four different attributes in our considerations and falls under one of the three quality ranges. A maximum clinical score of 4.5, 6.9 and 9.9 is taken as poor quality, standard quality and optimum quality respectively while scores less than 4.4 is ignored or render unsuitable for clinical measurement. All scores are normalised to range between (0 and 1) in our model.

### 2.5 Network Architecture

In the previous research [8], the multivariate network architecture was inspired by a couple of published works including Yang et al., [15] and Abdi et al.,[7]. Although using a single network to elicit feature extraction on each quality attributes was investigated, the performance recorded on each attribute vary significantly as network model show high variance to attributes data on chamber clarity and foreshortening. Echo images also presents varying complexities in the data structure hence was considered inadequate to extract features across multiple domain attributes efficiently. Consequently, a multi-stream architecture based on model subclassing was considered with implementation to accept full image size via transformation algorithm. Our model outperforms state-of-the-art models on all chosen attributes.

The multi-stream regression architecture consists of four parallel DCNN, each stream architecture incorporates slightly different component layers built specifically to generalise the different complexities of each quality attributes encountered by adapting a single network architecture. The four models were simultaneously trained on four attributes datasets. The combined streams architecture was then optimised to extract specific quality elements per frame independently. The architecture is depicted in Fig. 6., with respective details for each network stream. We defined a cost function using mean absolute error (MAE), the $l_1$ loss function via adaptive moment estimation (ADM) algorithm and simultaneously computes the model accuracy of each stream. The annotations, ground truth ($Q_{GT}$) of the videos were used as the quality score for all constituent frames of that video for model development and predicted score ($Q_P$) quality were evaluated to determine range of errors with respect to inter observer variability disparity and experts' score.

Each network in the multi-stream architecture accepts image input size of 227x227x3 as a 20 frames sequence and computes the weighted matrix of each frame pixel in order to perform discrete convolution which yields 2D output feature map ($F$) as stated in equation (7). The combine model yields four objective quality scores on each image/frame which represents the quality attributes as enumerated in 2.1 and 2.2. Of the four streams model architecture, three streams consist of four convolutional layers, interleaved with MaxPool (MP) and Batch Normalization (BN) layers, as depicted in Fig 6., were optimised for On-Axis, DepthGain and Foreshorten attributes while the fourth, consisting of three convolutional layers was specific to chamber clarity attributes. Each stream shares initial weights, defined in tensor flow API's loss weight's function and rectifier linear unit (ReLU) equation (5) was employed for inter-layer activation. While each stream yields a 2D features output, were flattened and are fed sequentially into a two layered, Long Short-Term Memory (LSTM) [16] to extract long term dependencies of images' temporal features. The final, linear layer provides the scores for each quality element through Sigmoid activation function defined in equation (6).

$$f(x)_{relu} = \max(0, x) \qquad (5)$$

$$f(x)_{sigmoid} = 1/\,(1 + e^{-x}{}_{i}) \qquad (6)$$

$$F^{l}_{(i,j)k} = \sum_{i=0}^{n}\sum_{j=0}^{m} w^{l}_{i,mn}\, F^{l-1}_{(j+m)(k+n)} \qquad (7)$$

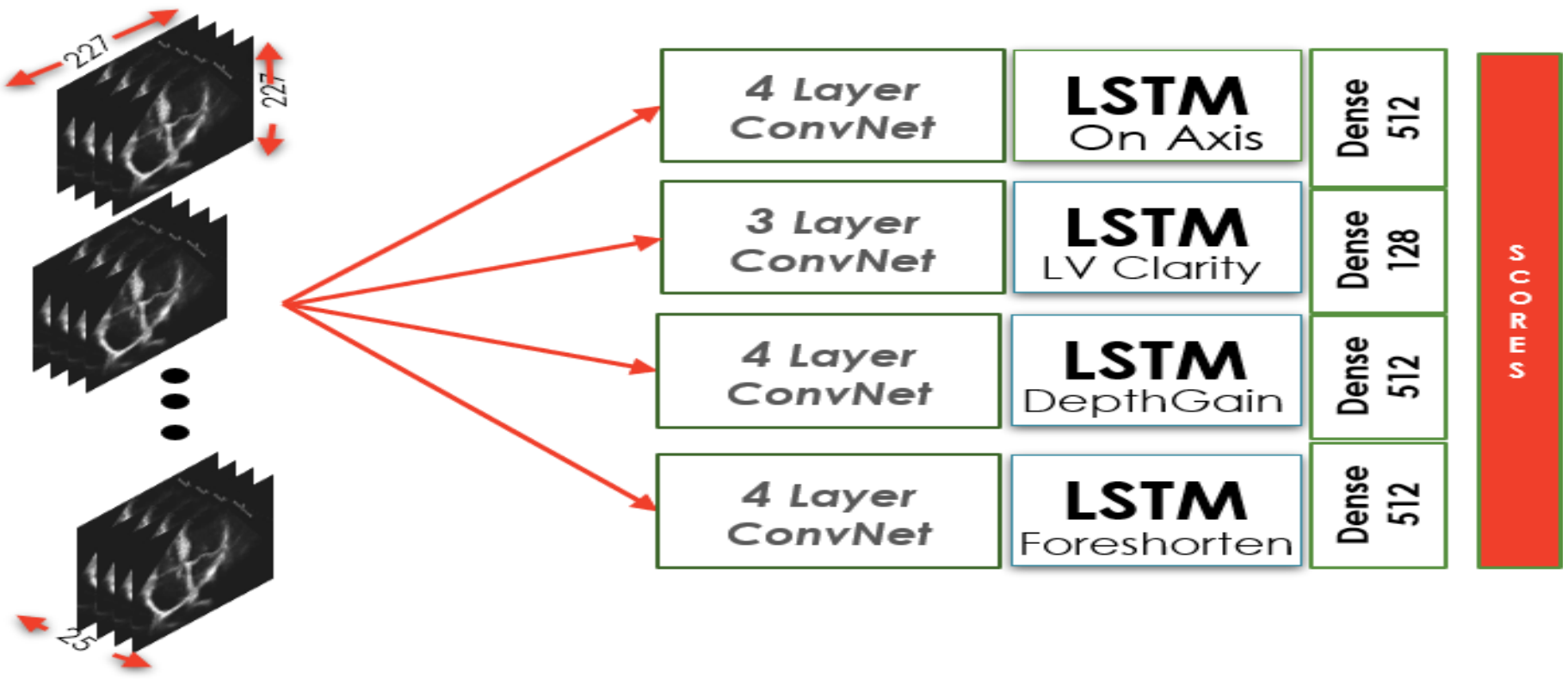

| Layers | On-Axis | LV Clarity | DepthGain | Foreshorten |
|---|---|---|---|---|
| 1st Layer | 32:3x3 | 16:3x3 | 32:3x3 | 16:3x3 |
| 2nd Layer | 32:3x3 | BN+MP | 32:3x3 | 16:3x3 |
| | MP | 32:3x3 | BN+MP | BN+MP |
| 3rd Layer | 64:3x3 | BN+MP | 64:3x3 | 32:3x3 |
| 4th Layer | 64:3x3 | 64:3x3 | 64:3x3 | 32:3x3 |
| | BN+MP | Dense128 | BN+MP | BN+MP |
| | Dense512 | | Dense512 | Dense512 |

Fig. 6. Multi-Stream network architecture used in the research showing numbers of kernels, corresponding sizes for each layer. Each stream is optimized to extract specific quality attributes and provides separate predicted scores on each quality attribute.

## 2.6 Training, Batch selection, Data augmentation

**Training hyper-parameters:** The model consists of four regression models arranged in parallel and were simultaneously trained using 5-fold cross validation technique to ensure adequate learning on the dataset and performance was recorded for each model. The hyper parameters learning rate was set at 0.0002 with high momentum 0.95 and decay rate of 0.1every 15 steps and were reproducibly initialized to minimise possible deviation in score performance. Training was initialised and completed as learning curves converged around 50 epochs.

**Batch selection:** The hardware computational cost during training phase ran high as batch selection of 8 and 12 were experimented, memory utilization becomes significantly apparent at batch selection of 12 at a fixed length sequence of 20 than running a batch size of 8 at the same fixed length sequence. Hardware performance difference of 0.13% in terms of computational speed was a negligible trade-off, did not affect the model's ability to properly generalize new test samples.

**Data augmentation:** Data augmentation was applied to allow optimum learning sequences for the models; a maximum translation of [-0.05, +0.05] pixels and maximum rotation of 10 degrees were applied randomly for horizontal, vertical and rotational angles, respectively. To prevent overfitting in the training phase, we applied batch normalization selectively, at specific convolution layer, early stopping and dropout (rate 0.32) for the training samples. Batch normalisation also helps stabilizes and speeds up convergence during the training phase.

**Hardware and software resources:** Model was implemented using TensorFlow backend. The experiment was carried out on a Z600 Intel i7 Quad Core mini server with 32GB memory and additional GPU GeForce GTX 970 chipset's Maxwell architecture and featuring 4GB RAM coupled to 1,664 CUDA cores.

## 2.7 Evaluation Metrics

Since the model uses multiplex variables for each score attributes, performance was evaluated via MAE and measured against absolute difference between cardiologist's score ($Q_{GT}$) and model's predicted automatic scores ($Q_P$) for each respective attributes and models, we computed model's class error in equation (8) as:

$$Class_{err} = \sum_{i=0}^{n} |Q_{GTi} - Q_{pi}| \qquad (8)$$

Minimal error, therefore, indicates best fit and better model performance. The average accuracy was computed in equation (9) as:

$$Model_{acc} = (1 - \sum_{i=0}^{n} |Q_{GTi} - Q_{pi}|) * 100 \qquad (9)$$

# 3 Results and Analysis

Given the complexity of varying pathological features in echo frames, our model could generalise on new echo frame with measured accuracy of 97.68 percent as shown in Table II. The error distribution per quality attributes is depicted in Fig. 7., for On-Axis, Clarity, DepthGain and Foreshortening attributes, respectively. The model prediction speed was found to be 0.012ms per frame for input pixel size of 227x 227 x 3, which is the assurance for real-time deployment and opportunity for enhancing clinical workflows.

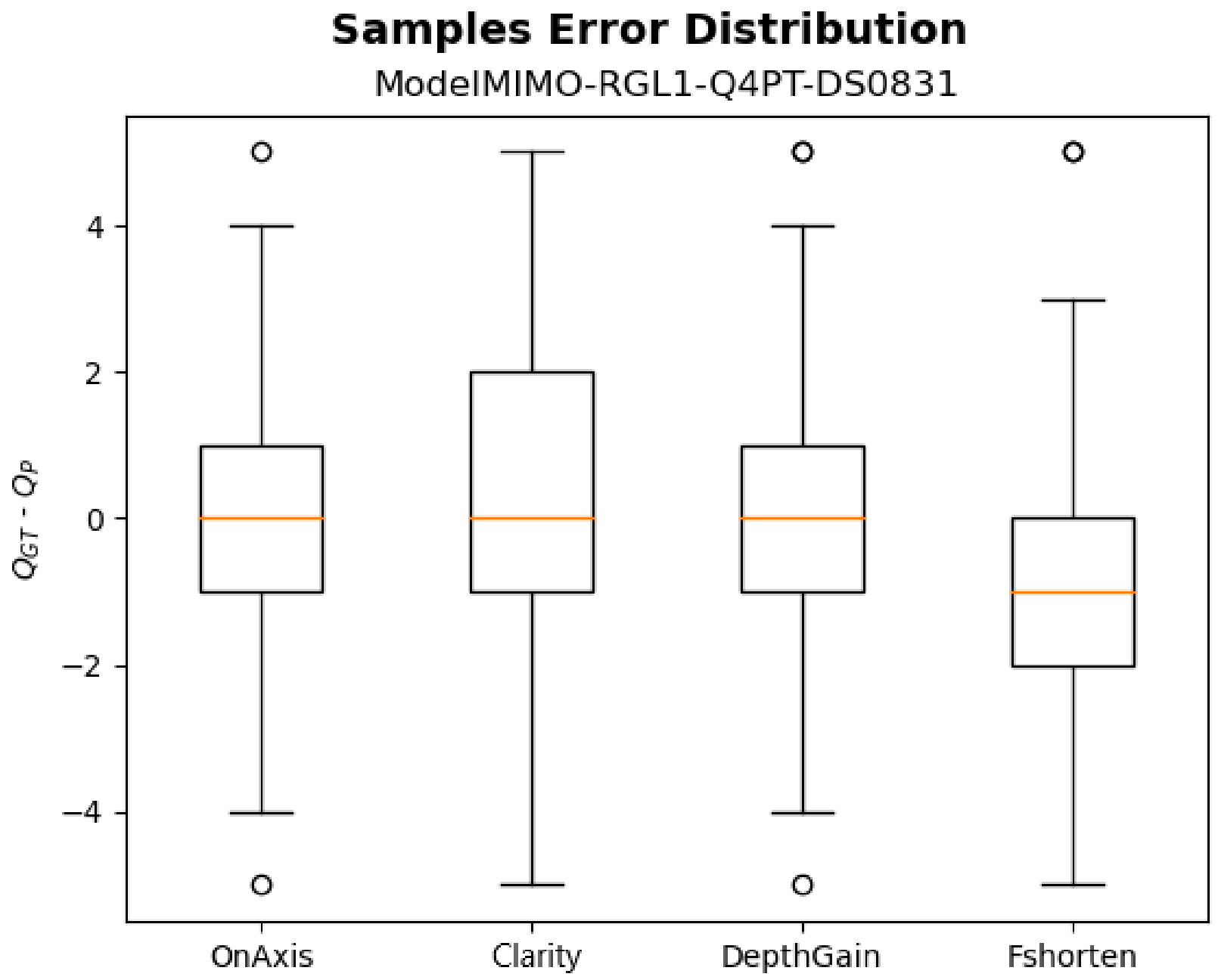

Fig. 7. Box Plot of the error distribution of test samples for each model. The error is computed as the difference between the prediction of the model and ground-truth. The x-axis shows four specific quality attributes of each model.

TABLE II
COMPUTED ACCURACY FOR THE TEST SAMPLES

| Models | On-Axis | LV Clarity | Depth Gain | Fore-Shortening | Average Accuracy |
|---|---|---|---|---|---|
| Model Accuracy | 97.69% | 97.51% | 97.81% | 97.70% | 97.68% |

### 3.1 Study limitation and future work

Unlike our previous study [8] where we consider only three quality attributes of A4C, this study is based on four quality attributes suitable for clinical image quality assessment and satisfying wider clinical requirements.

This proposed approach yielded a superior performance in terms of deployability and use case compared to any existing approach in quality assessment which indicates the specific element of image quality that must be optimised in real-time and provides objectivity on quality assessment score for operators' feedback and guidance.

Since this is a novel approach in quality assessment methods, a comparison with any existing approach would make a fair judgement. Unfortunately, existing assessment was only based on weighted average method of quality scoring did not provide any parallel approach to measure by. Hence, its near impossibility to determine equivalence while different dataset of corresponding annotation was utilised. Therefore, we make our dataset with expert annotations on four quality attributes public at IntSav repository to allow external validation by other researchers and equipment manufactures.

Similarly, we have considered A4C and A2C frames as the primary apical view standards to demonstrate the feasibility of clinical application for quality assessment, A2C quantifications may not be a focus under clinical practice suitable for unified quantification, therefore, future study would include other relevant apical view standards like PLAX, PSAX, A5C. again, several global characteristics can be used to distinguish between the different levels of quality and assignment of image quality index. Here, we only considered 4 attributes of image quality for clinical studies and research. A future possible study would include selective criteria that would be suitable for point of care deployment and encompass major laboratory assessments criteria.

Finally, we used two annotations or ground truth labels provided by two expert cardiologist and one accredited annotator. Intra-observer variability can be examined by obtaining additional annotations from human experts and compared with the error in the predicted scores.

## 4. CONCLUSION

We have presented the clinical significance and feasibility of developing an automated quality assessment in 2D echocardiographic images that contribute to automated diagnosis and quantification in echocardiology. An automated image quality assessment technique would be significant as part of a system that could accelerate the learning curve for those training in echocardiography and automated quality control process which is required for both clinical and research purposes. This would provide a real-time guidance to less experienced operators, increase their chances of acquiring optimum quality images and enhance diagnostic accuracy of cardiac functions.